\documentclass[letterpaper,10pt,twocolumn,final,journal,oneside]{IEEEtran}
\usepackage{array}
\usepackage{makecell}
\usepackage{epsfig}
\usepackage{graphics}
\usepackage{graphicx}
\usepackage[latin1]{inputenc}
\usepackage{amsmath}
\usepackage{amssymb}
\usepackage{booktabs}
\usepackage{subfigure}
\usepackage{bm}
\usepackage{cite}
\usepackage{cases}
\usepackage{color,soul}
\usepackage{comment}
\usepackage{stfloats}
\usepackage{multicol}
\usepackage{multirow}
\usepackage{balance}
\usepackage{acronym}
\usepackage{setspace}
\usepackage{cancel}
\usepackage{tikz}
\usepackage{hyperref}

\usepackage{siunitx}
\DeclareSIUnit{\atm}{atm}
\DeclareSIUnit{\kWh}{kWh}
\DeclareSIUnit{\Ah}{Ah}
\DeclareSIUnit{\OhmMeter}{\ohm\meter}
\DeclareSIUnit{\VA}{VA}

\acrodef{AFE}{active front end}
\acrodef{IMO}{International Maritime Organization}
\acrodef{AMP}{alternate marine power}
\acrodef{OPS}{on-shore power supply}
\acrodef{HVSC}{high voltage shore connection}
\acrodef{LRG}{low resistance grounding}
\acrodef{HRG}{high resistance grounding}
\acrodef{NGR}{neutral grounding resistor}

\IEEEoverridecommandlockouts

\newcommand\copyrighttext{%
  \footnotesize
  \centering\copyright~2021 IEEE. Personal use of this material is permitted. Permission from IEEE must be obtained for all other uses, in any current or future media, including reprinting/republishing this material for advertising or promotional purposes, creating new collective works, for resale or redistribution to servers or lists, or reuse of any copyrighted component of this work in other works.\\
  IEEE Transactions on Transportation Electrification.  \href{https://doi.org/10.1109/TTE.2021.3137717}{10.1109/TTE.2021.3137717}}
\newcommand\copyrightnotice{%
\begin{tikzpicture}[remember picture,overlay]
\node[anchor=south,yshift=0pt] at (current page.south) {\setlength{\fboxrule}{0pt}\fbox{\parbox{\dimexpr\textwidth-\fboxsep-\fboxrule\relax}{\copyrighttext}}};
\end{tikzpicture}%
}

\begin{document}

\title{High Voltage Shore Connection Systems: Grounding Resistance Selection and Short Circuit Currents Evaluation}

\author{Fabio~D'Agostino,~\IEEEmembership{Member,~IEEE,} Samuele~Grillo,~\IEEEmembership{Senior Member,~IEEE,} Roberto~Infantino, Enrico~Pons,~\IEEEmembership{Member,~IEEE}%
\thanks{F. D'Agostino is with the Dipartimento di Ingegneria Navale, Elettrica, Elettronica e delle Telecomunicazioni, Universit\`a degli Studi di Genova, Genova, Italy (e-mail: fabio.dagostino@unige.it).}%
\thanks{S. Grillo and R. Infantino are with the Dipartimento di Elettronica, Informazione e Bioingegneria, %
Politecnico di Milano, piazza Leonardo da Vinci, 32, I-20133 Milano, Italy %
(e-mail: samuele.grillo@polimi.it, roberto.infantino@mail.polimi.it).}%
\thanks{E. Pons is with the Dipartimento Energia ``Galileo Ferraris'', Politecnico di Torino, Torino, Italy (e-mail: enrico.pons@polito.it).}%
}

\IEEEaftertitletext{\copyrightnotice\vspace{0.2\baselineskip}}
\maketitle

\begin{figure*}[b]
{\centering
\includegraphics[width=\textwidth]{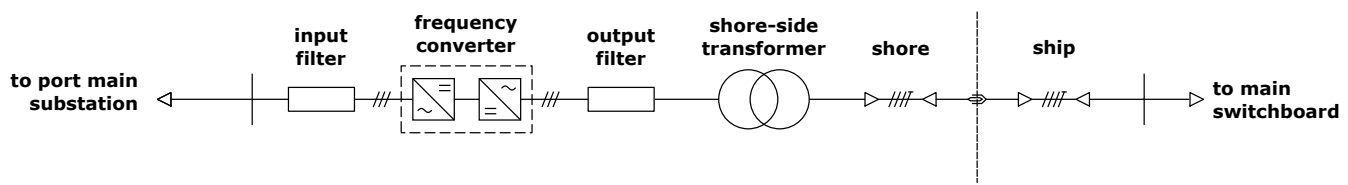}
\caption{Simplified single line diagram of the shore connection.}
\label{fig:DiagrammaUnifilare}}
\end{figure*}

\begin{abstract} Cold ironing represents an effective solution to remove air polluting emissions from ports. The high voltage shore connection system is the key enabling facility that allows to provide power from the shore side electrical system to the ship. The design of the shore connection needs a comprehensive assessment of the fault currents in different operating scenarios. International standards require the neutral point of the shore connection transformer be equipped with a neutral grounding resistor. Its value has to be defined to guarantee safety and protection of equipment and personnel in case of single phase-to-ground faults. Moreover, three-phase short circuits need to be considered to size equipment and protection devices. A crucial role is played by the frequency converter control system, required to adapt the mains frequency to the frequency of the ship. In this work, a complete electro-magnetic dynamic model of the high voltage shore connection has been developed, including frequency converter, shore-side transformer, connection MV cables and power system of the ship,  to analyze in detail the  behavior of the system in case of single phase-to-ground fault and three-phase short circuit, taking into account relevant standards and best practices.\end{abstract}



\acresetall
\section{Introduction}\label{sec:Introduction}

Maritime transport industry is recognized among the major sources of air pollution, especially in coastal areas~\cite{Pollution}. Following directives by \ac{IMO}~\cite{IMO2018}, EU~\cite{EU2019} and national ports' regulating authorities towards minimizing polluting emissions from ships, modern ports are going to play a key role by being transformed into smart grids~\cite{Kumar2019}, and smart energy hubs~\cite{john1}. In this context, design, installation and operation of cold ironing facilities, also referred to as shore connection, \ac{OPS} or \ac{AMP}, are becoming of paramount importance~\cite{ZisBook}.

The polluting emissions coming from auxiliary generators, required to supply essential services of docked ships, comprises carbon dioxide (${\rm CO}_2$), nitrogen oxides (${\rm NO_x}$), sulphur oxides (${\rm SO_x}$), and particulate matter (PM)~\cite{Stolz2021,Dago2021_2}. Cold ironing consists in supplying power to berthed ships by the shore side electrical system, allowing ships to switch off auxiliary engines. It can be considered as one of the most promising solutions to remove local air pollution from ports, when emission regulations will restrict or, even prohibit, the usage of diesel generators~\cite{ZisArticle}. Benefits are also expected to include noise pollution reduction. On the other hand one of the main challenges this transition is facing nowadays is represented by high investment costs~\cite{Europe}.

Shore connection facilities must be properly designed to meet operational and safety requirements, in order to satisfy the electrical load demand of different types of ship.  
In general, the cold ironing system comprises three subsystems: the shore-side power supply, the shore-to-ship interconnection, and the shipboard network. 
A simplified single line diagram of such configuration is shown in Fig.~\ref{fig:DiagrammaUnifilare}.

The first component is the port main substation which steps down the voltage to an acceptable level for the proper operation of the power electronics devices. Then, as the most common frequency value adopted on board is \SI{60}{\hertz}, while the electric grid in many parts of the world works at \SI{50}{\hertz}, a frequency conversion system is needed. The shore-side transformer located downstream the frequency converter has basically two functions: i) to adjust the voltage level reaching the value required by ship's main switchboard (\SI{6.6}{\kilo\volt} or \SI{11}{\kilo\volt}), and ii) to ensure that the line-to-ground fault at any location does not lead to any dangerous overvoltage conditions, that might cause safety issues and/or equipment damages. Finally, the shore-to-ship connection is constituted by cable reels, HV plug/socket-outlets with handling facilities, communication and control wires \cite{IEEE80005-2}. It is worth noting that, according to the Standards of Training, Certification, and Watch-keeping for Seafarers (STCW) Code, in marine applications the term high voltage refers to systems above \SI{1}{\kilo\volt}.

This paper focuses on the analysis of the \ac{HVSC} system when a single phase-to-ground fault or a three-phase short circuit occurs.

For what concerns single phase-to-ground faults, a thorough analysis about the shore power supply network structure and its grounding method is needed. In literature, extensive research has been carried out to determine which is the most adequate grounding method for the neutral point of the shore-side transformer \cite{Paul2006SystemCA, ACloserLook, Nelson, Paul2009ShoreToShipPS}. As a result, the grounding by means of a \ac{NGR}, also called neutral earthing resistor\footnote{The terms earthing and grounding are used interchangeably.}, has been chosen as the best compromise between the isolated and the solidly grounded systems \cite{Paul2006SystemCA}. However, it is possible to make a distinction between two different methods: \ac{LRG} and \ac{HRG}. The former consists in sizing the NGR in such a way that the resistive component of the fault current is significantly greater than the total system charging current, which is defined as the current that flows through the parasitic capacitances. The latter method implies that during a single phase-to-ground fault the resistive component of the fault current is slightly higher than the capacitive one. On the one hand, \ac{HRG}  provides lower damage risk at the fault location if compared to \ac{LRG}. For this reason, when the grounding resistance method was introduced, a high resistance was inserted in order to maintain the single phase-to-ground fault current below \SI{3}{\ampere} and to limit the arc flash hazard~\cite{Nelson}. On the other hand, LRG makes the tripping of the protection system easier, and implies lower values for transient overvoltages~\cite{Paul2009ShoreToShipPS}.

The second part of the analysis deals with the three-phase short circuit current, which is made up of two components: the contribution of the running on-board generators, and the contribution of the shore-side power system, which flows through the frequency converter. On board power stations are mainly composed by synchronous machines driven by diesel engines. Therefore the first component is well known in literature \cite{trifase,trifase2} and follows the behavior described in Standards IEC 61363-1 and IEC 60909-1 \cite{61363,60909}. The analysis of the behavior of the shore-side frequency converter is instead of higher interest.

The aim of the paper is twofold: i) to define a methodology to provide the range of admissible values for the neutral grounding resistor; and ii) to analyze the behavior of the frequency converter during a three-phase fault in order to give designing guidelines for the protection systems' sizing. To this end, we modeled the whole system to mimic the state-of-the-art condition, in terms of technology and control schemes.

The remainder of the paper is organized as follows.
In Section~\ref{sec:Problem definition} a short description of the current criteria related to \ac{HVSC} systems design is carried out in order to formulate the problem. In Section~\ref{sec:SystemModelling}, the shore power system, the shipboard power system and the grounding system models are presented. Single phase-to-ground fault scenarios are then discussed in Section~\ref{sec:Single phase-to-ground fault}. While in Section~\ref{sec:Three phase short circuit}, simulation results related to the three-phase short circuit case are analyzed. Finally, conclusions are drawn in Section~\ref{sec:Conclusion}.

\section{System design criteria} \label{sec:Problem definition}

When dealing with HV shore connection design, at least three standards should be taken into account: IEC/ISO/IEEE Standards 80005-1 \cite{IEEE80005-1}, IEC 61363-1 \cite{61363} and IEC 60909-0 \cite{60909}.

The first standard defines the main requirements for the connections between ships and shore power supplies. It points out the primary role of the NGR selection and recommends reference values for each kind of ship. However, it does not provide information about the effects of the NGR power rating on the fault current in the different operating scenarios.

The second standard presents procedures for calculating short-circuit currents on maritime power systems, but it does not provide information about cold ironing operation.

The third standard is used to calculate short circuit currents in three-phase ac systems and it is one of major references for terrestrial power systems. It includes a Section for the analysis of full size converters, but it refers to the values provided by the manufacturer and it does not provide information about the behavior of such components in case of fault.

The implementation of a highly detailed and complete model of the \ac{HVSC} system allows both to determine a range of values of the NGR compliant with the Standard requirements, and to investigate the behavior of the frequency converter in case of fault. 
The target system for the present study is a \SI{20}{\mega\VA} shore connection facility, suitable for cruise ships.


\subsection{System design practices}

As shown in Fig.~\ref{fig:DiagrammaUnifilare}, the shore-side transformer is connected  downstream the frequency converter.
According to IEC/ISO/IEEE 80005-1 \cite{IEEE80005-1} prescriptions, the transformer is equipped with a neutral earthing resistor on the secondary star-connected side. The earthing terminal of this \acs{NGR} is connected to the ship hull through the neutral cable, and to the ground by a neutral earthing-switch operated depending on the type of ship.
Thus, the single phase-to-ground fault current is mainly determined by the value of such resistor, and the steady-state value of the ground fault current is expected to be very low, if compared to the nominal value of phase currents. Moreover, in order to avoid any potential hazards to personnel, Standard and best practices impose the following constraints:
\begin{enumerate}
    \item An earth fault shall not create a step or touch voltage exceeding \SI{30}{\volt} at any location in the shore-to-ship power system~\cite{IEEE80005-1}.
    \item The resistive component of the fault current shall not be less than 1.25 times the prospective system charging current~\cite{Designing}.
\end{enumerate}
It is worth pointing out that when studying phase-to-ground faults on vessels, one usually refers to the case in which a phase conductor touches an exposed conductive part, which is bonded to the metallic hull of the ship. Therefore, to avoid the presence of dangerous touch voltages at any location of the shore connection facility, an equipotential bonding between ship's hull and shore earthing system is prescribed. The first of the two aforementioned requirements is intended to investigate this situation.

On the other hand, to meet the second requirement, the total system charging current must be compared with the total resistive current. This aspect is crucial for the design of the grounding system. Hence, in order to size the NGR, both the components of the fault current must be known. However, an accurate total system charging current evaluation is not always possible during the ship-to-shore infrastructure design stage. Moreover, the value of the ship-side contribution could vary substantially from one ship to another. Therefore, a good practice can be to size the NGR case by case. However in the everyday life it is not possible to change the value of the NGR according to the parameters of the docked ship. This paper intends to investigate this aspect providing a range of admissible values for the NGR.
\vspace{9pt}

\subsection{Cable System and Bonding Requirements}\label{sec:BondingRequirements}
A shore facility, suitable for the connection of cruise ships, utilizes four power cables with phase conductors laid up with an earthing core. The four earthing cores, whose cross-section has to be at least 50\% of the power cores cross-section, realize the equipotential bonding among ship's hull and shore earthing system.
A neutral single pole cable provides a solid bonding between the earthing terminal of shore power transformer's NGR and ship's hull. In addition, an earthing switch (or disconnect switch \cite{Designing}) connects the NGR's earthing terminal to the shore grounding system. Indeed, Standard \cite{IEEE80005-1} requires that, during cruise ship operation, the neutral earthing resistor is connected only to the ship-side by the neutral cable, while it has to be connected to the ground when a ship is not connected. This aspect will be further discussed in Section~ \ref{sec:Single phase-to-ground fault}.

\subsection{Fault analysis}
IEC 61363-1 and IEC 60909-0 \cite{61363,60909} are the two major standards for the calculation of the short circuit currents. The former, which is dedicated to electrical installations of ships, identifies the procedure to calculate three-phase short circuit current as a time-dependent function due to the presence of the sub-transient and transient modes of synchronous generators taking into account the effect of external line impedance on the time constant \cite{trifase}. The latter allows to calculate fault currents based on the well-known equivalent voltage source method. However, when full size converters are considered, they shall be modelled in the positive-sequence system by a current source whose value has to be provided by the manufacturer (Section 6.9 of  Standard IEC 60909-0).

One of the goals of this paper is to investigate further this aspect showing the transient behavior of the output currents of the frequency converter during three-phase short circuits.

\section{System modeling} \label{sec:SystemModelling}

In this Section, the different components of an HVSC system at \SI{11}{\kilo\volt} are described in detail. The complete model has been implemented in the Matlab Simulink simulation environment.

\subsection{Shore-side system design}

\begin{figure}[b]
{\centering
\includegraphics[width=\columnwidth]{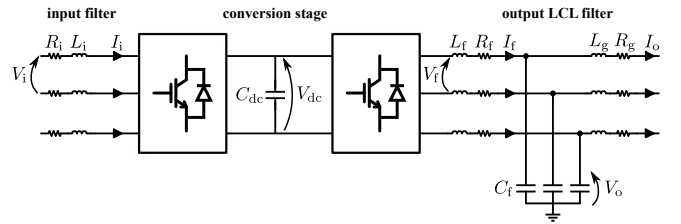}
\caption{Schematic diagram of the frequency converter. Resistances in input and output filters model the resistances of the inductances.}
\label{fig:FreqConv}}
\end{figure}

The first and main component of the shore power supply system is the frequency converter. It is made up of two stages: an ac-dc and a dc-ac converter. For what concerns the former stage, the use of diode rectifiers typically represents the cheapest solution. However, the utilization of an \ac{AFE} converter, with controllable switches, leads to numerous benefits, such as: higher power factor on the grid side, controllability of the dc-link voltage, sinusoidal input currents, smaller dc link voltage ripple (leading to a corresponding dc capacitor size reduction) \cite{Mohan}. Moreover, this technology allows bidirectional flow of current and it is suitable for reversible power applications, where the ship can provide power and services to the distribution grid \cite{Integration}. For these reasons, an IGBT based switch-mode rectifier has been selected.

\begin{figure*}[t]
  \centering
  \begin{tabular}{c}
  \subfigure[]{\label{fig:fig_3_1}\includegraphics[width=\textwidth]{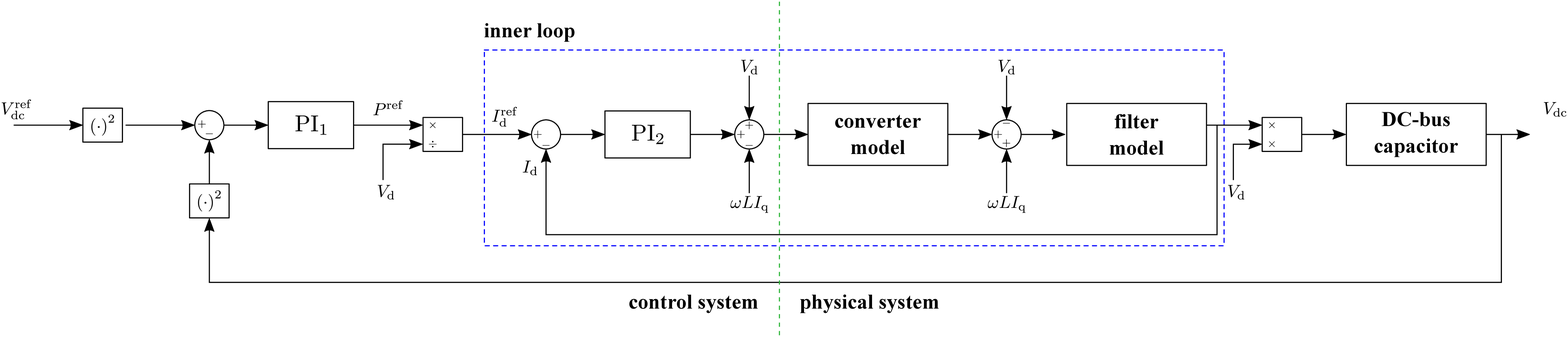}}\\
  \subfigure[]{\label{fig:fig_3_2}\includegraphics[width=.55\textwidth]{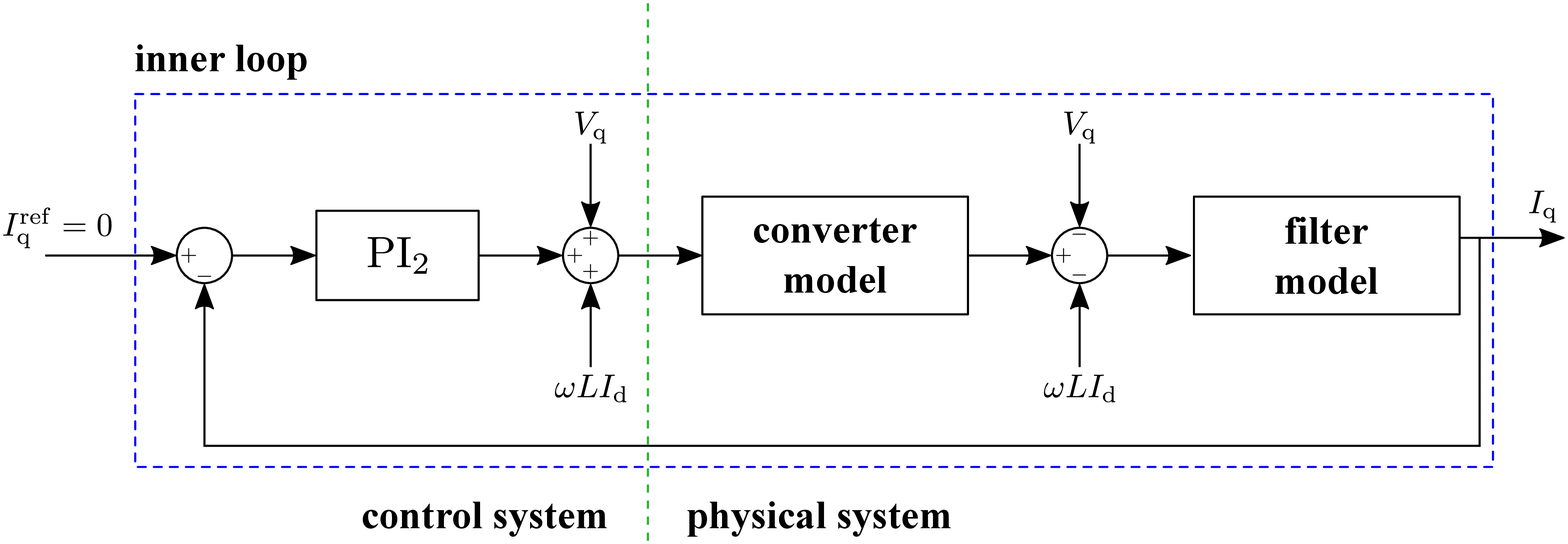}}
  \end{tabular}
\caption{Complete scheme of the control system of the rectifier in a dq reference frame: (a) d-axis; (b) q-axis.}\label{fig:ControlSystemRectifier}
\end{figure*}

\begin{figure*}[t]
{\centering
\includegraphics[width=\textwidth]{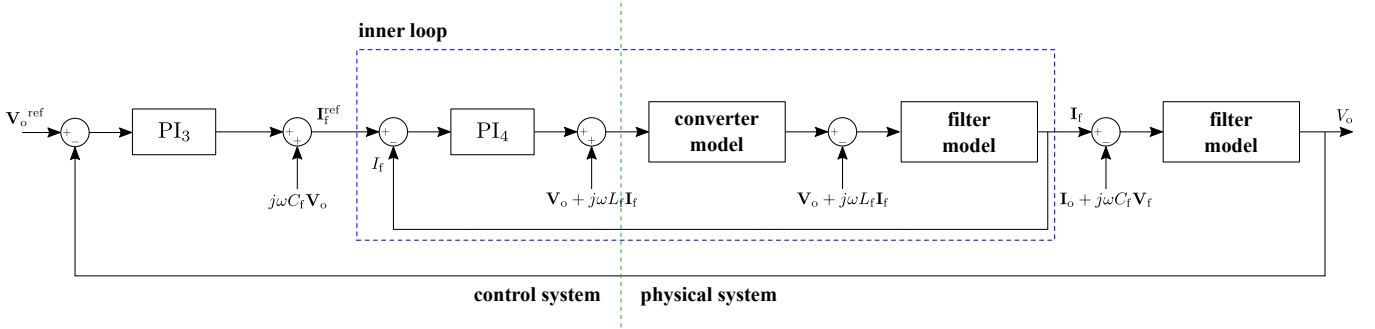}
\caption{Complete scheme of the control system of the inverter. All the variables are expressed as Park vectors in a dq reference frame.}
\label{fig:ControlSystemInverter}}
\end{figure*}

Figure~\ref{fig:FreqConv} shows the scheme of the frequency converter that has been implemented in the present work. Since the output of each converter is resulting from a PWM modulation, the adoption of filters is required for reducing as much as possible the harmonic content. The values of such filters are reported in Table~\ref{tab:inverter}.

To achieve a proper operation of the converter the control method adopted in this paper is based on a double loop. It consists of an inner loop for controlling the current and an outer loop for controlling the voltage \cite{Perini}. In the case of the ac-dc conversion, such control method aims at regulating the dc output voltage (Fig.~\ref{fig:ControlSystemRectifier}). Thus, the converter works to maintain the dc voltage at a desired reference value, using a feedback PQ control. In this case, the comparison is performed between the square of the measured voltage and the square of the reference value. In this way the tuning of the controller is easier, as the square of the dc voltage is proportional to the active power exchanged with the dc bus~\cite{Grillo:2014}.

The control method of the inverter (Fig.~\ref{fig:ControlSystemInverter}) aims at keeping constant the voltage across the capacitor of the output filter $\left(C_{\rm f}\right)$. All the parameters of the control systems are reported in Table~\ref{tab:ControlloConverter}.

\begin{table}[ht]
\small
    \centering
    \caption{Parameters of the Frequency Converter.}
    \label{tab:inverter}
\begin{tabular}{l S[table-format=2.2]}
 \toprule
 \multicolumn{2}{c}{Parameters of the rectifier}\\
 \midrule
Input ac voltage $(V_{\rm i})$ & \SI{15}{\kilo\volt}\\
Nominal output dc voltage $(V_{\rm dc})$& \SI{28}{\kilo\volt}\\
minimum output dc voltage $\left( V^{\rm min}_{\rm dc} = \frac{2\sqrt{2}V_{\rm i}}{\sqrt{3}} \right)$ & \SI{24.49}{\kilo\volt}\\
Input filter resistance $(R_{\rm i})$  & \SI{0.25}{\ohm} \\
Input filter inductance $(L_{\rm i})$  & \SI{5.26}{\milli\henry} \\
DC bus capacitor  $(C_{\rm dc})$  & \SI{5}{\milli\farad} \\
\toprule
\multicolumn{2}{c}{Parameters of the inverter} \\
\midrule
Output ac voltage $(V_{\rm f})$ & \SI{15}{\kilo\volt}\\
Output filter first stage resistance $(R_{\rm f})$  & \SI{0.12}{\ohm}  \\
Output filter first stage inductance  $(L_{\rm f})$  & \SI{2.39}{\milli\henry} \\
Output filter capacitance $(C_{\rm f})$  & \SI{0.1}{\milli\farad} \\
Output filter second stage resistance  $(R_{\rm g})$  & \SI{0.81}{\ohm} \\
Output filter second stage inductance $(L_{\rm g})$  &  \SI{1.28}{\milli\henry} \\
\bottomrule
\end{tabular}
\end{table}

\begin{table}[ht]
\small
    \centering
    \caption{Parameters of the Control Systems.}
    \label{tab:ControlloConverter}
\begin{tabular}{l l S}
 \toprule
 \multicolumn{3}{c}{Parameters of the rectifier control system} \\
 \midrule
\multirow{4}{*}{${\rm PI}_{1}$ controller} & integral gain & \SI{12}{\second^{-1}}\\
& proportional gain & \SI{0.48}{}\\
& cut-off frequency  & \SI{100}{\hertz} \\
&  phase margin  & \SI{75}{\degree} \\
\hline
\multirow{4}{*}{${\rm PI}_{2}$ controller}  & integral gain & \SI{6291}{\second^{-1}}\\
& controller: proportional gain & \SI{17}{} \\
& controller: cut-off frequency  & \SI{1000}{\hertz} \\
& controller: phase margin  & \SI{75}{\degree} \\
\toprule
\multicolumn{3}{c}{Parameters of the inverter control system} \\
\midrule
\multirow{4}{*}{${\rm PI}_{3}$ controller}  & integral gain & \SI{1.1}{\second^{-1}}\\
& proportional gain & \SI{0.019}{} \\
& cut-off frequency  & \SI{200}{\hertz} \\
& phase margin  & \SI{75}{\degree} \\
\hline
\multirow{4}{*}{${\rm PI}_{4}$ controller}  & integral gain & \SI{16837}{\second^{-1}}\\
& proportional gain & \SI{47}{}\\
& cut-off frequency  & \SI{2000}{\hertz} \\
& phase margin  & \SI{75}{\degree} \\
 \bottomrule
\end{tabular}
\end{table}

\subsection{Ship-side system model design}

One of the common arrangements of primary generators on board modern vessels consists in synchronous machines driven by diesel prime movers \cite{Aleksandar}. The dynamic model of each generating unit is formed by the following components: a synchronous machine, a prime mover, a diesel engine speed governor (GOV), and an excitation system, which includes the automatic voltage regulator (AVR) \cite{DP3}.

The GOV model includes a speed control transfer function, an actuator transfer function, and the combustion delay of the prime mover. For on-board power systems, the adoption of the droop-based control mode for the active power and frequency control is typically recommended. For this reason, the well-known DEGOV1 model, developed for Woodward governors, has been selected.

The excitation system model is based on the IEEE Standard 421.5 \cite{Recommended}. The model selected for this study is AC5C.

\subsection{Ship load model design}

In the present study, a \SI{15}{\mega\VA} electrical load has been modelled. During cold ironing operation, it is assumed that the loads connected in a cruise ship are mainly hotel loads. In this paper a ZIP model has been used for the loads. It is a composition of the three types of constant components (Z: constant impedance; I: constant current; P: constant power). The key assumption behind the selection of the ZIP parameters is that when the ship is berthed the behavior of hotel loads can be assimilated to residential loads. However, in the case of a modern cruise ship, it is reasonable to expect a consistent use of air conditioning, which is controlled by inverters~\cite{Guerrero}. As a consequence, the constant power contribution is relevant~\cite{p_load}. The values that have been adopted in this work are ${\rm Z}=60\%$; ${\rm I}=15\%$; ${\rm P}=25\%$.

\subsection{Cable connection model design} \label{sec:Cables}
As previously described in Section~\ref{sec:BondingRequirements}, an intentional equipotential bonding between the ship's hull and the ground collector busbar shall be realized by means of a specific core in the power cables. In many works in literature such wire is treated as a zero impedance connection \cite{Designing,ACloserLook,Kozak}. However, as this study is focused on the evaluation of touch voltage, it is necessary to take into account the real impedance of such connection, $Z_{\rm E}$. In addition to the bonding conductor, a neutral conductor connects the NGR to the ship's hull, with an impedance $Z_{\rm N}$, and when the NGR disconnect switch is closed, the two conductors are in parallel.

The connection between the shore side and the ship is realized according to the Standard's prescriptions, as described in Section~\ref{sec:BondingRequirements}. The parameters adopted for the study have been selected from a real data sheet of MV shielded cables. Both cable formation and parameters are reported in Table~\ref{tab:cables}.

In the preliminary simulations that were carried out, it was noticed that the presence of inductive and parasitic capacitive parameters of neutral and earth bonding conductors was highly increasing the computational time even if their effect was negligible on the final numerical results. As a matter of fact, small variations in the neutral and earthing conductor parameters do not affect the results significantly, given that the neutral grounding resistor and ship's earthing resistance play a major role in limiting the single-phase-to-ground fault current. Moreover, such small variations could also be produced by different cables' lengths. For this reason, we decided to neglect the inductive and capacitive parameters of both neutral and bonding conductors. Thus, these conductors have been modeled as purely resistive components. The values of the ``equivalent'' resistances have been set to be the magnitude of the conductors' impedances, i.e., $R_{\rm E}=$ \SI{5.7}{\milli\ohm} and $R_{\rm N}=$ \SI{9.2}{\milli\ohm}. On the other hand, the full model of the phases' cores allows to catch the most significant behavior of the system during three-phase short-circuits, when the earthing conductors are not involved.

\begin{figure*}[t]
{\centering
\includegraphics[width=.9\textwidth]{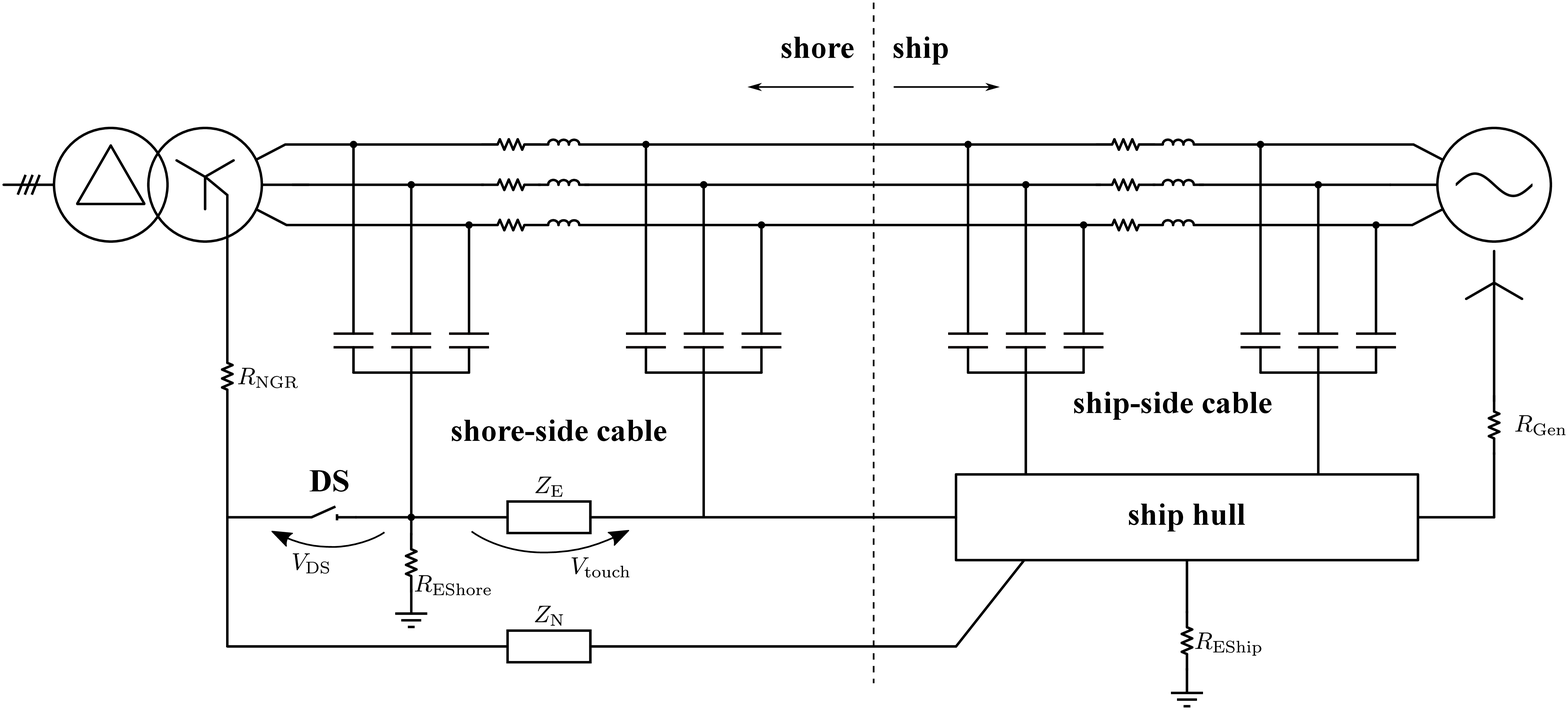}
\caption{Complete scheme of grounding and cable systems of the shore connection.}
\label{fig:grounding}}
\end{figure*}

As mentioned in Section~\ref{sec:Introduction}, one of the aims of this paper is to investigate the effects of different operating conditions on touch voltage. This quantity is usually defined as the voltage drop across the bonding conductors, as depicted in Fig.~\ref{fig:grounding}~\cite{ACloserLook}. We performed simulations with disconnect switch open and closed and we measured the touch voltage, $V_{\rm touch}$, and the electric potential difference across the terminals of the disconnect switch, $V_{\rm DS}$, in each simulation.

\subsection{Grounding system model design} \label{sec:GroundingSystem}

The equipotential bonding is earthed at both sides: on the shore and on the ship. Therefore, the overall model, shown in Fig.~\ref{fig:grounding}, must include also these resistances: $R_{\rm EShore}$ and $R_{\rm EShip}$. The former is the earth resistance of the shore earthing system. Generally, it is possible to exploit the port grounding system. By doing so, the resulting resistance is very small (e.g., \SIrange[range-phrase = --]{0.4}{1.2}{\ohm}) \cite{Examination}. The latter is the earth resistance of the ship. Its value, which is variable for each ship, shall be measured case by case.
Key parameters for the estimation of such resistance are the hull's immersed surface, hull painting and salty water resistivity \cite{Shore2Ship}. For a cruise ship, it is reasonable to assume a surface equal to \SI{10.000}{\meter^2} \cite{Examination}. The other two quantities are very complex to be estimated. It is possible to find both new vessels with the hull perfectly insulated by a high-resistivity coating (\SI{e9}{\OhmMeter}) and old vessels with the hull practically bare (resistivity in the range of \SI{e2}{\OhmMeter}) \cite{Examination}. For what concerns water resistivity, the two main variables are temperature and salinity. The computation of the resistivity has been carried out using the Debye-Huckel-Osnager equation \cite{Equation}. It results in values around \SI{0.2}{\OhmMeter} for oceans, while less salty seas can present values in the range of \SIrange[range-phrase=--]{1}{2}{\OhmMeter}. As a result, $R_{\rm EShip}$ can vary from \SI{0.3}{\ohm} to \SI{4}{\ohm}.

Before starting with the fault analysis, it is necessary to consider also the grounding system of the ship. The focus is posed on the grounding arrangement of the neutral point of the diesel generator. In almost all cases related to cruise ships at \SI{11}{\kilo\volt}, the neutral point is grounded through a high-value current limiting resistor ($R_{\rm gen}$) \cite{DNVGL48}, designed to respect the maximum single-phase-to-ground fault current limit of \SI{5}{\ampere} \cite{Islam2018}.

In order to use realistic values for the model parameters, we performed an extensive literature review and analyzed the schematics of existing shore-ship interconnections. Then, for cable parameters we used values taken from the data sheet of cable manufacturers. For other important parameters, such as $R_{\rm EShip}$ we resorted to information found in literature that we further elaborated. The parameters used are summarized in Table~\ref{tab:cables}.

\begin{table}[ht]
\small
    \centering
    \caption{Parameters of Cables and Grounding System.}
    \label{tab:cables}
\begin{tabular}{l S[table-alignment=right]}
\toprule
\multicolumn{2}{c}{Multipolar 3 phases + earth bonding cable} \\
\midrule
Phase cross section & \SI{240}{\milli\meter\squared}\\
Phase resistance at \SI{65}{\celsius} & \SI{0.0927}{\ohm\per\kilo\meter}\\
Phase reactance at \SI{60}{\hertz} & \SI{0.1163}{\ohm\per\kilo\meter} \\
Phase capacitance  & \SI{0.423}{\micro\farad\per\kilo\meter} \\
Earth core cross section & \SI{120}{\milli\meter\squared}\\
Earth core resistance at \SI{65}{\celsius} & \SI{0.182}{\ohm\per\kilo\meter}\\
Earth core reactance at \SI{60}{\hertz} & \SI{0.1356}{\ohm\per\kilo\meter} \\
Cable formation & {\(4 \times (3 \times 240 + 1 \times 120)\)}\\
\toprule
\multicolumn{2}{c}{Unipolar neutral cable} \\
\midrule
Cross section & \SI{240}{\milli\meter\squared}\\
Resistance at \SI{65}{\celsius} & \SI{0.0920}{\ohm\per\kilo\meter}\\
Reactance at \SI{60}{\hertz} & \SI{0.1236}{\ohm\per\kilo\meter} \\
Cable formation & {\(1 \times 240\)}\\
\toprule
\multicolumn{2}{c}{Parameters of the grounding system} \\
\midrule
\makecell[l]{Earth resistance of the\\ shore earthing system, $R_{\rm EShore}$} & \SI{1}{\ohm}\\
Earth resistance of the ship, $R_{\rm EShip}$ & \SI{1}{\ohm}\\
\bottomrule
\end{tabular}
\end{table}

To sum up, the whole system that has been implemented is shown in Fig.~\ref{fig:grounding}. It includes the shore-side substation and its grounding resistance ($R_{\rm NGR}$), the diesel generator and its grounding resistance ($R_{\rm gen}$), shore-side and ship-side cables, represented as lumped $\pi$-model equivalent circuit, the equipotential bonding conductors ($R_{\rm E}$), the neutral conductor ($R_{\rm N}$), the NGR grounding disconnector  and the shore and ship grounding systems' resistances ($R_{\rm EShore}$ and $R_{\rm EShip}$).

\section{Single phase-to-ground fault} \label{sec:Single phase-to-ground fault}

In this Section, the simulation results related to the single phase-to-ground fault are presented.
As summarized in Table~\ref{tab:scenarios}, the simulations include four different scenarios:
\begin{itemize}
\item \emph{Scenario 1 (SC1)}: HVSC has been just activated, i.e., the synchronization process has been completed and the ship's network is connected to the shore's one. However, the diesel generator (DG) is still in operation, and it is still feeding the whole ship's load ($P_{\rm load}$).
\item \emph{Scenario 2 (SC2)}: shore connection is activated, and the ship's load is supplied from the shore side. The power set point of the diesel generator ($P_{\rm DG}$) is null, as a consequence of the load transfer. However, the generator grounding system is still active.
\item \emph{Scenario 3 (SC3)}: shore connection is activated. The diesel generator and its grounding system are disconnected. The load is entirely supplied by shore power ($P_{\rm HVSC}$).
\item \emph{Scenario 4 (SC4)}: shore connection is activated, and it is absorbing power from the ship. The ship's load is lower than in the previous cases, and the diesel generator is providing services to the terrestrial network. The on-board grounding system is active.
\end{itemize}
For each scenario, eight values of the neutral grounding resistor of the shore-side transformer $R_{\rm NGR}$, are analyzed. For the present study, a zero-impedance fault has been considered. In addition to this, for each scenario and for each value of the NGR two operating conditions have been simulated: i) with the disconnect switch open; and ii) with the disconnect switch closed.

\begin{table}[ht]
\small
    \centering
    \renewcommand{\arraystretch}{1.3}
    \caption{Definition of the Simulation Scenarios.}
    \label{tab:scenarios}
\begin{tabular}{l|cccc}
\toprule
                                   & SC1   & SC2   & SC3   & SC4   \\
\midrule
DG           & on    & on    & off   & on    \\
DG grounding & on    & on    & off   & on    \\
$P_{\rm DG}$     & \SI{15}{\mega\volt\!\ampere} & 0     & ---     & \SI{15}{\mega\volt\!\ampere} \\
$P_{\rm load}$   & \SI{15}{\mega\volt\!\ampere} & \SI{15}{\mega\volt\!\ampere} & \SI{15}{\mega\volt\!\ampere} & \SI{10}{\mega\volt\!\ampere} \\
$P_{\rm HVSC}$   & 0     & \SI{15}{\mega\volt\!\ampere} & \SI{15}{\mega\volt\!\ampere} & \SI{-5}{\mega\volt\!\ampere} \\
\bottomrule
\end{tabular}
\end{table}

\begin{figure}[ht]
{\centering
\includegraphics[width=.9\columnwidth]{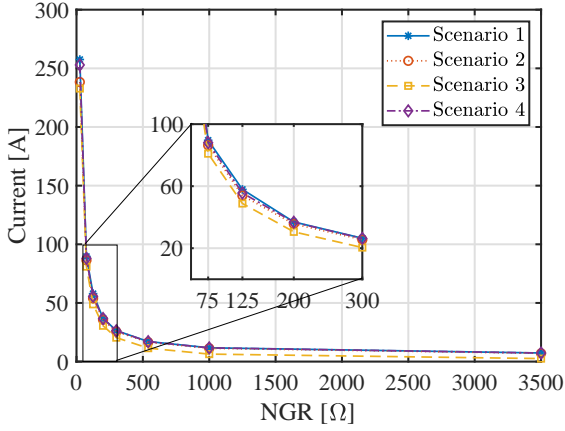}
\caption{Comparison of the steady-state rms single-phase-to-ground fault currents in the different scenarios. The disconnect switch is open. The curves for the four scenarios with disconnect switch closed are practically identical, and are here omitted.}
\label{fig:Ifault_quad}}
\end{figure}

In Fig.~\ref{fig:Ifault_quad}, a comparison of the rms values of the steady-state fault current for the four scenarios with disconnect switch open is reported. The cases with disconnect switch closed are not shown, as the values of the fault current are practically identical.  The first thing to notice is that the absence of the contribution from the generator's grounding implies a lower fault current in the third scenario with respect to the others.

\begin{figure}[ht]
{\centering
\includegraphics[width=.9\columnwidth]{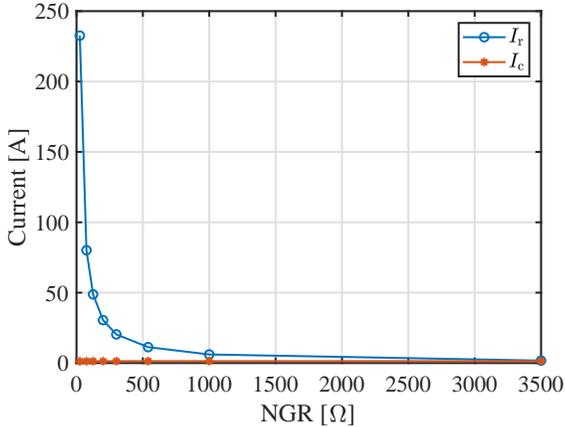}
\caption{Scenario no. 3: comparison between the resistive ($I_{\rm r}$) and capacitive ($I_{\rm c}$) components of the fault current.  The disconnect switch is open. The curves for the corresponding scenario with disconnect switch closed are practically identical---as it happens in all other scenarios---, and are here omitted.}
\label{fig:I3_quad}}
\end{figure}

\begin{table*}[t]
\small
    \centering
    \renewcommand{\arraystretch}{1.3}
    \caption{Power to Be Dissipated by the NGR in Each Scenario and for Each Different Value of the NGR.}
    \label{tab:power}

\begin{tabular}{l|SSSSSSSS}
\toprule
 & \SI{25}{\ohm}   & \SI{75}{\ohm}  & \SI{125}{\ohm}  & \SI{200}{\ohm}  & \SI{300}{\ohm} & \SI{540}{\ohm} & \SI{1}{\kilo\ohm} & \SI{3.5}{\kilo\ohm}\\
\midrule
SC1 & \SI{1591.2}{\kilo\watt} & \SI{533.2}{\kilo\watt} & \SI{320.2}{\kilo\watt} & \SI{200.2}{\kilo\watt} & \SI{133.5}{\kilo\watt} & \SI{74.2}{\kilo\watt} & \SI{40.1}{\kilo\watt} & \SI{11.5}{\kilo\watt}\\
SC2 & \SI{1361.9}{\kilo\watt} & \SI{482.7}{\kilo\watt} & \SI{298.7}{\kilo\watt} & \SI{187.2}{\kilo\watt} & \SI{125.3}{\kilo\watt} & \SI{70.0}{\kilo\watt} & \SI{39.0}{\kilo\watt} & \SI{10.7}{\kilo\watt}\\
SC3 & \SI{1352.6}{\kilo\watt} & \SI{481.7}{\kilo\watt} & \SI{297.7}{\kilo\watt} & \SI{186.2}{\kilo\watt} & \SI{124.2}{\kilo\watt} & \SI{69.0}{\kilo\watt} & \SI{37.2}{\kilo\watt} & \SI{10.6}{\kilo\watt}\\
SC4 & \SI{1558.8}{\kilo\watt} & \SI{526.2}{\kilo\watt} & \SI{316.4}{\kilo\watt} & \SI{198.0}{\kilo\watt} & \SI{132.1}{\kilo\watt} & \SI{73.4}{\kilo\watt} & \SI{39.6}{\kilo\watt} & \SI{11.3}{\kilo\watt}\\
\bottomrule
\end{tabular}
\end{table*}

In Fig.~\ref{fig:I3_quad}, the resistive component of the fault current $\left(I_{\rm r}\right)$ in Scenario 3 is compared with the total system charging current $\left(I_{\rm c}\right)$. What emerges is that in the eighth case study $\left(R_{\rm NGR}=\SI{3500}{\ohm}\right)$, the two curves are very close. In particular $I_{\rm r}=\SI{1.745}{\ampere}$ and $I_{\rm c}=\SI{1.370}{\ampere}$. As a consequence, it can be concluded that Scenario 3 imposes the upper limit of the NGR to \SI{3500}{\ohm}. Indeed, if a higher value were to be selected, the resistive current would be lower than 1.25 times the system charging current. This would violate the second condition reported in Section~\ref{sec:Problem definition} (Par. 6.2.3 \cite{IEEE80005-1}).

On the other hand, Scenario 1 imposes the lower limit for the NGR (to \SI{125}{\ohm}). As clearly shown in Fig.~\ref{fig:Ifault_quad}, if a lower value were to be selected, the fault current would be higher than \SI{60}{\ampere}, a value generally considered as undesirable \cite{Paul2006SystemCA}.

\begin{figure}[ht]
\centering
\includegraphics[width=.9\columnwidth]{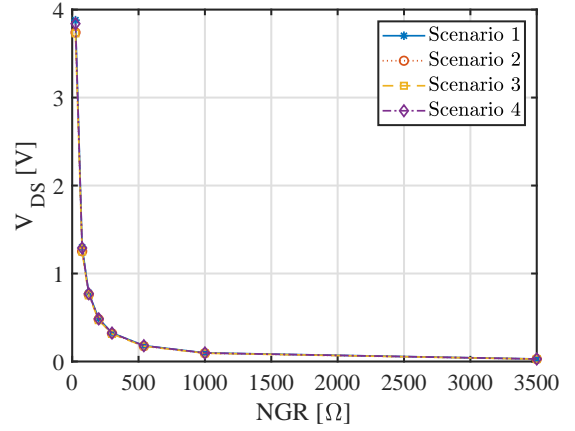}
\caption{$V_{\rm DS}$ when the disconnect switch is open in the different scenarios.}
\label{fig:touch_1}
\end{figure}

\begin{figure}[ht]
\centering
\includegraphics[width=.9\columnwidth]{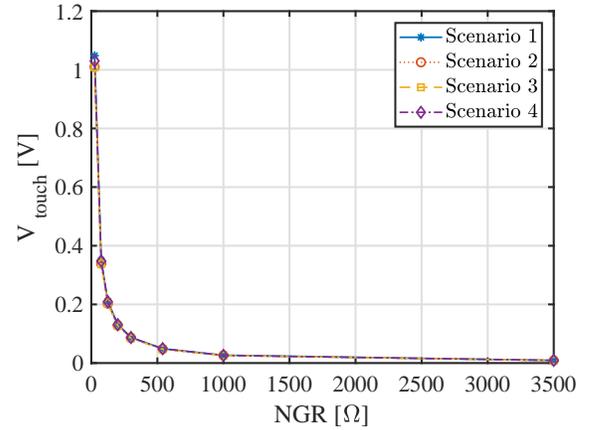}
\caption{$V_{\rm touch}$ when the disconnect switch is closed in the different scenarios.}
\label{fig:touch_2}
\end{figure}

In order to assess the touch voltage limit condition, we performed simulations with disconnect switch open and closed, and we measured the touch voltage, $V_{\rm touch}$, in each simulation. We found that when the disconnect switch is open this values is negligible (being it in the order of magnitude of \si{\milli\volt}, thus justifying this practice for cruise ships). For this reason we further investigated the issue, and monitored the electric potential difference across the terminals of the disconnect switch, $V_{\rm DS}$, as reported in Fig.~\ref{fig:touch_1}. Figure~\ref{fig:touch_2} shows the steady-state rms values of $V_{\rm touch}$ when the disconnect switch is closed. In both cases, the limit threshold of \SI{30}{\volt} is never exceeded.

Finally, Table~\ref{tab:power} reports the required power rating of the NGR to continuously withstand the fault current. For the acceptable values of the NGR the power is between \SI{10}{\kilo\watt} and \SI{320}{\kilo\watt}. Those values are compatible with existing commercial equipment.

\section{Three-phase short circuit} \label{sec:Three phase short circuit}

This Section is focused on the analysis of the three-phase short circuit at the main switchboard of the ship. The contribution of the on-board generator can be evaluated in a quite straightforward way with the application of Standards IEC 60909~\cite{60909} and IEC 61363~\cite{61363}. For this reason, this issue is not analyzed in detail in this work.

On the contrary, a methodology for calculating the contribution of the frequency converter to the short-circuit current is not provided by the Standard, which only states that full size converters can be modelled in the positive-sequence system by a current source, whose value has to be provided by the manufacturer. For this reason, in this Section the contribution of the shore-side frequency converter is analyzed with the objective of defining a set of equations to analytically calculate it, as it is normally done for the contribution of synchronous generators.

\begin{figure}[ht]
{\centering
\includegraphics[width=.9\columnwidth]{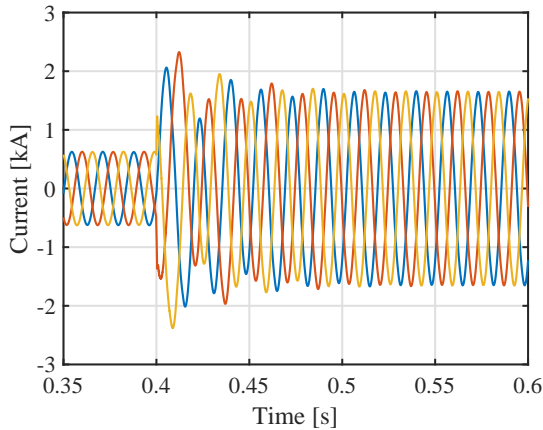}
\caption{Contribution of the frequency converter $i(t)$ to the three-phase short-circuit current.}
\label{fig:I_trifase}}
\end{figure}

Using the model implemented in Simulink, dynamic simulations have been performed. The results are shown in Fig.~\ref{fig:I_trifase}.
It can be observed that in the first hundreds of milliseconds after the fault event the total output current, $i^{\rm ph}(t)$, is made up of two components: a symmetrical component at \SI{60}{\hertz}, $i^{\rm ph}_{\rm s}(t)$, and a decay oscillating component, ${i_{\rm dec}(t)}$:
\begin{equation}
  i^{\rm ph}(t)= i^{\rm ph}_{\rm s}(t)+i_{\rm dec}(t).
  \label{Iph}
\end{equation}
In order to analyze the factors that affect the trend of the latter component, a moving average at \SI{60}{\hertz} was applied. By doing so, the symmetrical component, which is known and equal to the pre-fault voltage divided by the commutation reactance of the converter, has been removed:
\begin{equation}
  i^{\rm ph}_{\rm s}(t)=I_{\rm s} \sin \left ( 2\pi 60 t+\phi^{ph} \right )
  \label{Is}
\end{equation}
leaving only the decaying aperiodic function:
\begin{equation}
  i_{\rm dec}(t)= a_{2} \exp\left(-b_{2} t\right) \sin \left ( 2\pi f_{2} t \right ).
  \label{Idec}
\end{equation}
The coefficients of \eqref{Idec} are reported in Table~\ref{tab:coefficients2}.
\begin{table}[ht]
\small
    \centering
    \renewcommand{\arraystretch}{1.3}
    \caption{Coefficients of Equation~\eqref{Idec} When the Integral Gain of ${\rm PI_1}$, ${\rm K_{I}}$, and the Saturation Limit of ${\rm PI_4}$ are Modified. When the Integral Gain Varies, the Saturation is set to $\pm\SI{10}{\kilo\volt}$, When the Saturation varies, the Integral Gain is set to $\SI{12}{\per\second}$.}
    \label{tab:coefficients2}
\begin{tabular}{l S[table-number-alignment = right] S[table-number-alignment = right] S[table-number-alignment = right]}
\toprule
& \multicolumn{1}{c}{$a_{2}$} & \multicolumn{1}{c}{$b_{2}$} & \multicolumn{1}{c}{$f_{2}$}\\
\midrule
${\rm K_{I}}=\SI{12}{\per\second}$ & \SI{804}{\ampere} & \SI{27}{\per\second} & \SI{22}{\hertz}\\
${\rm K_{I}}= \SI{20}{\per\second}$ & \SI{801}{\ampere} & \SI{40}{\per\second} & \SI{22}{\hertz}\\
${\rm K_{I}}= \SI{25}{\per\second}$ & \SI{798}{\ampere} & \SI{54}{\per\second} & \SI{22}{\hertz}\\
${\rm K_{I}}= \SI{30}{\per\second}$ & \SI{792}{\ampere} & \SI{60}{\per\second} &  \SI{22}{\hertz}\\
\midrule
${\rm Sat}= \pm\SI{10}{\kilo\volt}$  & \SI{804}{\ampere} & \SI{27}{\per\second} & \SI{22}{\hertz}\\
${\rm Sat}= \pm\SI{12}{\kilo\volt}$ & \SI{805}{\ampere} & \SI{28}{\per\second} &  \SI{25}{\hertz}\\
${\rm Sat}= \pm\SI{15}{\kilo\volt}$ & \SI{805}{\ampere}& \SI{28}{\per\second} &  \SI{28}{\hertz}\\
${\rm Sat}= \pm\SI{20}{\kilo\volt}$ & \SI{811}{\ampere}& \SI{29}{\per\second} & \SI{32}{\hertz}\\
\bottomrule
\end{tabular}
\end{table}
Fitting tools have been exploited to derive the equations of the curves, and a relation between ${i_{\rm dec}}$ and the transient behavior of the voltage at the dc link ${v_{\rm dc}}$ has been identified. It can be observed that the voltage at the dc link can be expressed as:
\begin{equation}
  v_{\rm dc}(t)= a_{1} \exp\left(-b_{1} t\right) \left ( 1+c_{1} \sin \left( 2\pi f_{1} t\right ) \right ) + V_{\rm dc_{0}},
  \label{Vdc}
\end{equation}
where:
\begin{equation}
  V_{\rm dc_{0}}= \SI{28}{\kilo\volt} = 1 {\rm p.u.}
\end{equation}

\begin{figure}[ht]
{\centering
\includegraphics[width=.9\columnwidth]{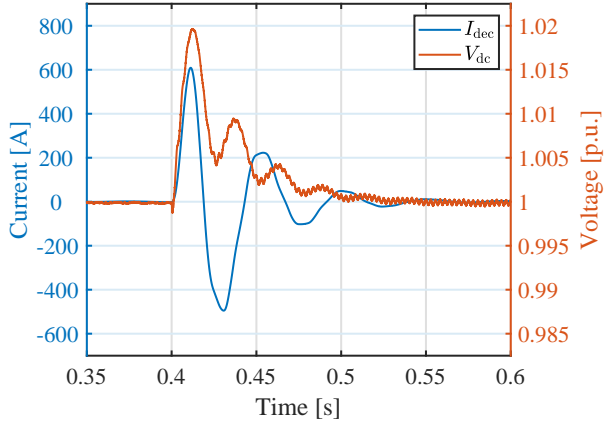}
\caption{Decaying aperiodic component $i_{\rm dec}(t)$ of the frequency converter contribution $i(t)$ and voltage at the dc link $v_{\rm dc}(t)$.}
\label{fig:I_dec}}
\end{figure}

Figure~\ref{fig:I_dec} shows that the decaying time of the voltage at the dc link $\tau_{\rm dc}=1/b_{1}$ is equal to the time constant of the decaying aperiodic component of the current $\tau_{\rm i}=1/b_{2}$.

\begin{figure}[ht]
{\centering
\includegraphics[width=.9\columnwidth]{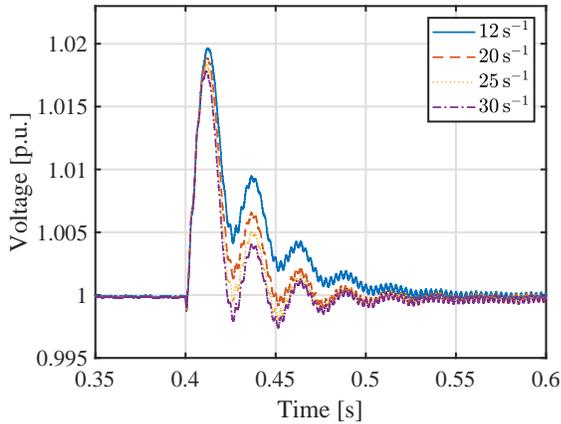}
\caption{Behavior of the voltage at the dc link $v_{\rm dc}(t)$ by varying the integral gain ${\rm K_{I}}$ of the ${\rm PI_{1}}$ controller. The value of the saturation of the controller ${\rm PI_{4}}$ is set to $\pm\SI{10}{\kilo\volt}$.}
\label{fig:Vdc_trifase}}
\end{figure}

\begin{figure}[ht]
{\centering
\includegraphics[width=.9\columnwidth]{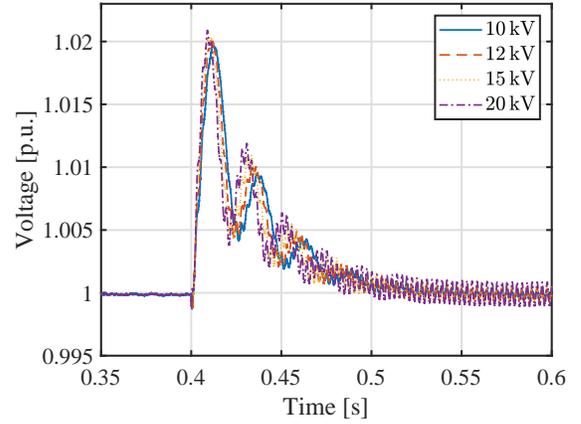}
\caption{Behavior of the voltage at the dc link $v_{\rm dc}(t)$ by varying the saturation limit of the ${\rm PI_{4}}$ controller. The value of the integral gain of the controller ${\rm PI_{1}}$ is set to $\SI{12}{\per\second}$.}
\label{fig:Vdc_f}}
\end{figure}

\begin{table}[ht]
\small
    \centering
    \renewcommand{\arraystretch}{1.3}
    \caption{Coefficients of Equation~\eqref{Vdc} When the Integral Gain of ${\rm PI_1}$, ${\rm K_{I}}$, and the Saturation Limit of ${\rm PI_4}$ are Modified. When the Integral Gain Varies, the Saturation is set to $\pm\SI{10}{\kilo\volt}$, When the Saturation varies, the Integral Gain is set to $\SI{12}{\per\second}$.}
    \label{tab:coefficients}
\begin{tabular}{l S S S S}
\toprule
& \multicolumn{1}{c}{$a_{1}$} & \multicolumn{1}{c}{$b_{1}$} & {$c_{1}$} & \multicolumn{1}{c}{$f_{1}$}\\
\midrule
${\rm K_{I}}=\SI{12}{\per\second}$ & \SI{410}{\volt} & \SI{27}{\per\second}& \num{0.55} & \SI{38}{\hertz}\\
${\rm K_{I}}= \SI{20}{\per\second}$ & \SI{395}{\volt} & \SI{40}{\per\second}& \num{0.61} & \SI{38}{\hertz}\\
${\rm K_{I}}= \SI{25}{\per\second}$ & \SI{395}{\volt} & \SI{54}{\per\second}& \num{0.68} & \SI{38}{\hertz}\\
${\rm K_{I}}= \SI{30}{\per\second}$ & \SI{380}{\volt} & \SI{60}{\per\second}& \num{0.74} & \SI{38}{\hertz}\\
\midrule
${\rm Sat}= \pm\SI{10}{\kilo\volt}$ & \SI{410}{\volt} & \SI{27}{\per\second}& \num{0.55} & \SI{38}{\hertz}\\
${\rm Sat}= \pm\SI{12}{\kilo\volt}$ & \SI{420}{\volt} & \SI{28}{\per\second}& \num{0.55} & \SI{42}{\hertz}\\
${\rm Sat}= \pm\SI{15}{\kilo\volt}$ & \SI{423}{\volt} & \SI{28}{\per\second}& \num{0.54} & \SI{46}{\hertz}\\
${\rm Sat}= \pm\SI{20}{\kilo\volt}$ & \SI{426}{\volt} & \SI{29}{\per\second}& \num{0.53} & \SI{50}{\hertz}\\
\bottomrule
\end{tabular}
\end{table}

As reported in Table~\ref{tab:coefficients}, the simulation empirically proves that $\tau_{\rm dc}=1/b_{1}$ depends on the coefficients of the outer loop regulator of the control system of the rectifier (${\rm PI_{1}}$ shown in Fig.~\ref{fig:ControlSystemRectifier}). Particularly, increasing the integral gain ${\rm K_{I}}$, the time constant decreases, resulting in a faster transient (Fig.~\ref{fig:Vdc_trifase}). On the other hand, the frequency $f_{1}$ depends on the choice of the saturation limit of the inverter's inner control loop (${\rm PI_{4}}$ shown in Fig.~\ref{fig:ControlSystemInverter}). Particularly, increasing the saturation limit, the frequency increases (Fig.~\ref{fig:Vdc_f}).

\section{Conclusion}\label{sec:Conclusion}

This work analyzes in detail the different components of high voltage shore connection systems for modern ships. The relevant standards have been discussed and the system behavior in case of single phase-to-ground fault and three-phase short circuit has been analyzed. For this purpose, a complete electro-magnetic dynamic model has been developed in Matlab Simulink, and several scenarios have been simulated, including the special case when the ship delivers power to the terrestrial distribution network. In order to use reasonable values for the model parameters, an extensive literature review was performed, existing shore-ship interconnections schematics were analyzed and manufacturers' data sheets were used.

In the case of single phase-to-ground fault, a crucial role is played by the value of the neutral grounding resistor of the shore-side transformer, while the operation with earthing switch open results in lower values of the touch voltage. The results show that the admissible values of the NGR range between \SI{125}{\ohm} and \SI{3500}{\ohm}, where ``admissible'' means that the rules imposed by reference Standards and best practice are not violated in any operating scenario.

On the other hand, in the case of three-phase short circuit, a crucial role is played by the outer loop regulator of the rectifier and by the inner current control loop of the inverter. In fact, there is a strict relation between the voltage at the dc link and the decay oscillating component of the short circuit current, which impacts the peak short-circuit current. In particular, thanks to the simulation results, empirical equations have been obtained to describe the impact of control system parameters on the short circuit current of the frequency converter.




\end{document}